\documentclass[twocolumn]{aastex631} 

\usepackage{graphics,epsf}
\usepackage[utf8]{inputenc}
\usepackage{amsmath}                
\usepackage{amsfonts}               
\usepackage{amssymb}                
\usepackage{epsfig}                 
\usepackage{graphicx}               
\usepackage{float}
\usepackage{color}
\usepackage{multirow}               
\usepackage{hyperref}

\hypersetup{
    colorlinks=true,
    linkcolor=red,   
    urlcolor=cyan}

\usepackage[colorinlistoftodos]{todonotes}

\newcommand{\s}{{~\rm s}}

\newcommand{\g}{{~\rm g}}
\newcommand{\G}{{~\rm G}}
\newcommand{\K}{{~\rm K}}

\newcommand{\kpc}{{~\rm kpc}}

\begin{document}

\title{Identifying a point-symmetrical morphology in the core-collapse supernova remnant W44}

\author[0000-0003-0375-8987]{Noam Soker}
\affiliation{Department of Physics, Technion, Haifa, 3200003, Israel; soker@physics.technion.ac.il}

\begin{abstract}
I identify a point-symmetrical morphology in the core-collapse supernova remnant (CCSNR) W44 compatible with shaping by three or more pairs of jets in the jittering jet explosion mechanism (JJEM). Motivated by recent identifications of point-symmetrical morphologies in CCSNRs and their match to the JJEM, I revisit the morphological classification of CCSNR W44. I examine a radio map of W44 and find the outer bright rim of the radio map to possess a point-symmetric structure compatible with shaping by two energetic pairs of opposite jets rather than an S-shaped morphology shaped by a precessing pair of jets. An inner pair of filaments might hint at a third powerful pair of jets. More pairs of jets were involved in the explosion process. This study adds to the growing evidence that the JJEM is the primary explosion mechanism of core-collapse supernovae.  
\end{abstract}

\section{Introduction}
\label{sec:Introduction}

The jittering jets explosion mechanism (JJEM) is a theoretical mechanism for exploding core-collapse supernovae (CCSNe), in which several to a few tens of pairs of jets explode the star. 
Intermittent accretion disks around the newly born neutron star (NS) launch the exploding jets in stochastically varying directions and powers (e.g., \citealt{PapishSoker2011, GilkisSoker2014, ShishkinSoker2021, WangShishkinSoker2024}; for recent quantitative values of the different parameters of the JJEM, see \citealt{Soker2024Learning}). 
Most jets are choked inside the core, deposit their energy, and explode the star. Some pairs of jets might leave imprints on the CCSN remnant (CCSNR), e.g., by forming dense clumps, inflating bubbles or ears, breaking out through the main ejecta, and leaving behind a nozzle.

Since the intermittent accretion disks launch pairs of opposite jets, these structural features come in pairs. Pairs of structural features not along the same line form a point-symmetric morphology. As such, the JJEM predicts that many CCSNRs possess point-symmetric morphologies, but not all CCSNRs.
The recent identification of point-symmetric morphologies in more than ten CCSNRs confirms one of the fundamental predictions of the JJEM (for a review of 2024 results and list of the point-symmetric CCSNe see \citealt{Soker2025Two}). 

The effect of the NS natal kick, instabilities during the explosion process, and the interaction of the ejecta with a circumstellar material and the interstellar medium cannot account for all properties of the identified point-symmetric morphologies of these CCSNRs (\citealt{SokerShishkin2024Vela}). These processes smear any point-symmetrical morphology. These make it challenging to identify point-symmetrical morphologies in some cases.
For these challenging identifications, any additional CCSNR with an identified point-symmetric morphology is a treasure to JJEM researchers. 

As I show in this study, the CCSNR W44 (G34.7-0.4, 3C 392) is also challenging. The earlier W44 morphological classification was of an elongated structure with an S-shape morphology (Section \ref{sec:Shape}). Motivated by the recent identification of point-symmetrical CCSNRs, I revisit its morphological classification and find it to be point-symmetric (Section \ref{sec:Point}).  
In Section \ref{sec:Counter}, I discuss alternatives that cannot account for the point-symmetric morphology, and in Section \ref{sec:Summary}, I summarize this short study and its significant contribution in establishing the JJEM as the primary explosion mechanism of CCSNe.  

\section{Previously S-shape classification}
\label{sec:Shape}

In this section, I discuss the previous morphological classification of W44 concerning a jet-driven explosion.  

\begin{figure}
\begin{center}
\includegraphics[trim=0cm 11.0cm 2cm 0.0cm,scale=0.52]{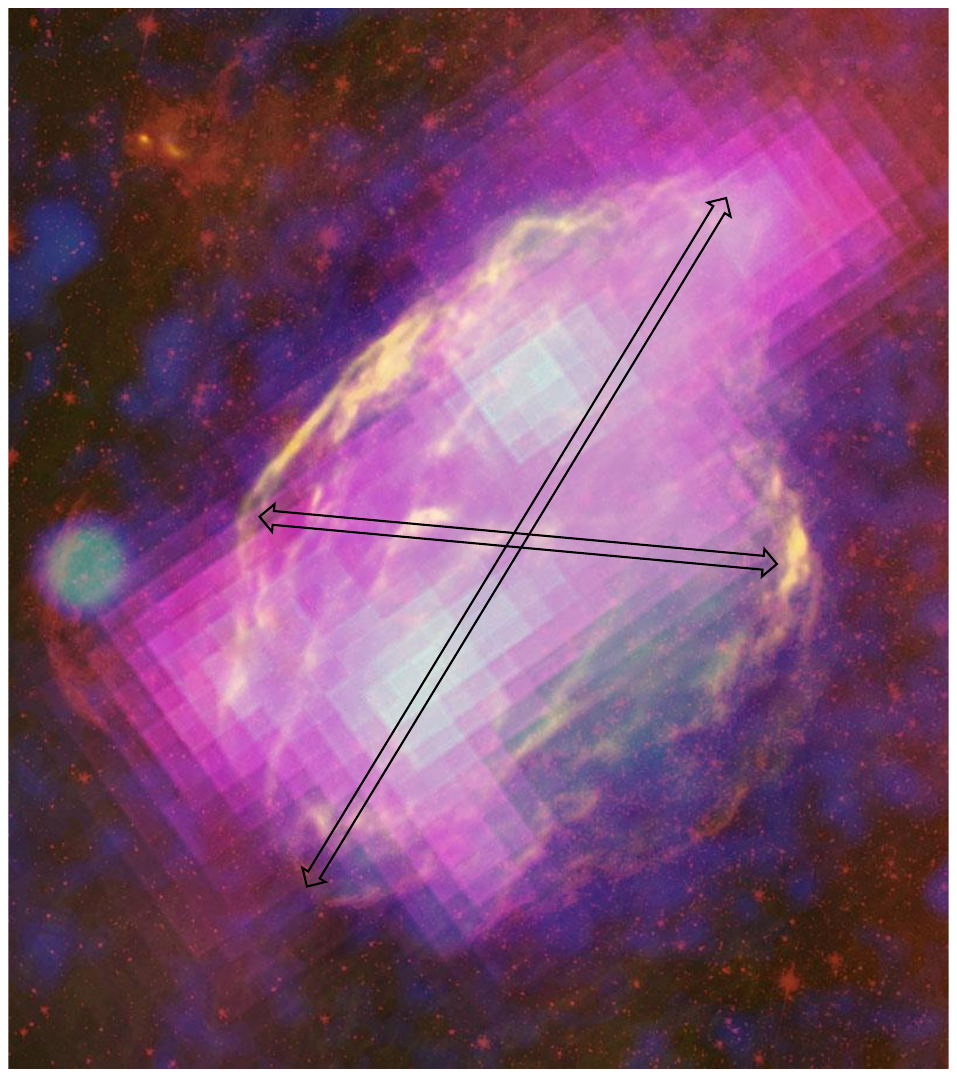} 
\caption{A composite image of CCSNR W44 from NASA site. I added two double-headed arrows that depict the two axes of the two energetic jets that I propose to have shaped W44 during the explosion process (Section \ref{sec:Point}). 
Magenta: GeV gamma-ray from Fermi’s LAT; yellow: Radio from the Karl G. Jansky Very Large Array; Red: infrared; Blue: X-ray from ROSAT. 
Credit: NASA/DOE/Fermi LAT Collaboration, NRAO/AUI, JPL-Caltech, ROSAT; \tiny{\url{https://www.nasa.gov/universe/nasas-fermi-proves-supernova-remnants-produce-cosmic-rays/}}
}
\label{fig:colored}
\end{center}
\end{figure}

SNR W44 was mapped in several wavebands, e.g., Gamma-ray (e.g., \citealt{Abdoetal2010}),  X-ray (e.g.\citealt{Jonesetal1993, Sheltonetal2004, Kawasakietal2005, Uchidaetal2012, Okonetal2020}), visible (e.g., \citealt{Rhoetal1994, Giacanietal1997, Mavromatakisetal2003}), IR (e.g., \citealt{Reachetal2005, Reachetal2006}) and radio (e.g., \citealt{Setraetal2004, Hoffmanetal2005, Castellettietal2007, Anderletal2014, Egron2017, Loruetal2019}).  SNR W44  is close to the Galactic plane and interacts with a molecular cloud (e.g., \citealt{Rhoetal1994, Setaetal1998, Nobukawaetal2018, Kooetal2020, Liuetal2022, Cosentinoetal2023}). It has a pulsar, PSR B1853+01 and its distance is $\simeq 3 \kpc$ (e.g., \citealt{Wolszczanetal1991}).  
In Figure \ref{fig:colored}, I present a composite image of W44 from the NASA site.
The image shows the complicated structure of W44, particularly its filamentary texture and unequal west-east sides. I mark the two axes of the two energetic jets that I propose shaped W44 during the explosion process (Section \ref{sec:Point}). 

\begin{figure}[t]
\begin{center}
\includegraphics[trim=4cm 14.0cm 1cm 2.5cm,scale=0.6]{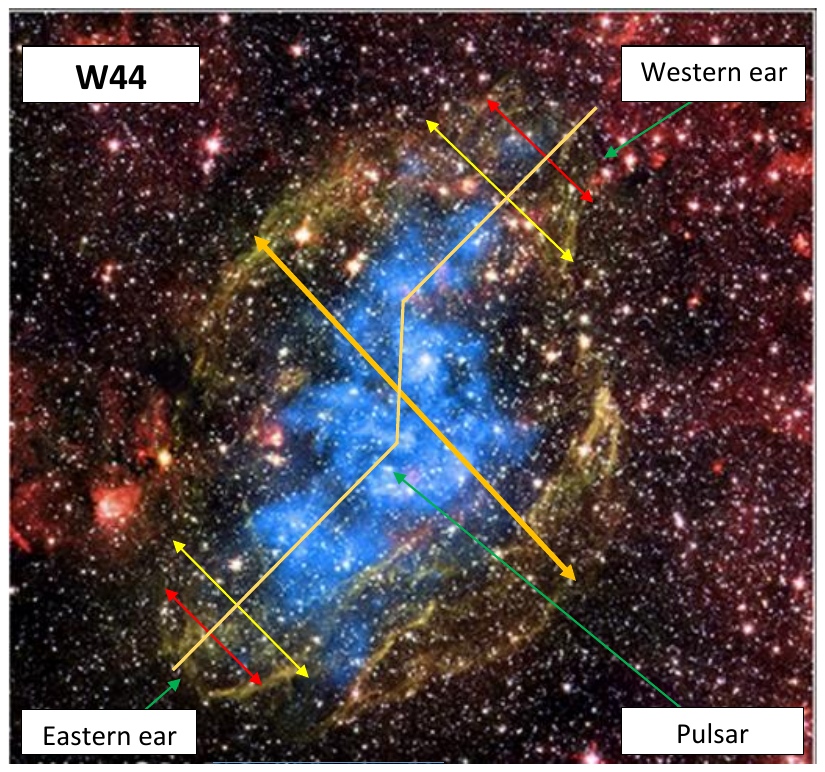} 
\caption{An image of CCSNR W44 from the Chandra gallery, with marks from \citet{GrichenerSoker2017}. Red, blue
and green colors represent infrared emission (based on NASA/JPL-Caltech; see image in \citealt{Reachetal2006}), and cyan represents X-ray emission (based on \citealt{Sheltonetal2004}).
The red and yellow double-headed arrows mark the two opposite ears that \citet{GrichenerSoker2017} identify, and the three-segment line is the S-shape that they attributed to SNR W44. 
}
\label{fig:W44GrichenerSoker}
\end{center}
\end{figure}

W44 has an elongated morphology (e.g., \citealt{Castellettietal2007}). \citet{GrichenerSoker2017} identified two ears in IR images of W44. Figure \ref{fig:W44GrichenerSoker} is from their paper.
They mark the base of each ear that they identified (thin double-headed yellow arrows), the middle of each ear (thin double-headed red arrows), and the center of W44 in the symmetrical morphology they identified (thick orange double-headed arrow).     
They aimed to calculate the energy of the jets that inflated the ears.   

\cite{GrichenerSoker2017} identified an S-shaped morphology, as the three-segment line on Figure \ref{fig:W44GrichenerSoker} depicts. They also note that the ears they define with the S-shaped morphology appear in the IR image but not in the X-ray or radio. Motivated by identifying eleven point-symmetric CCSNRs in the past two years, I revisit the morphological classification of W44.   

\section{Identifying a point-symmetric morphology in W44}
\label{sec:Point}
Considering W44's complicated structure, I prefer to examine the radio map, which is the most detailed. I determine whether W44 or part of it possesses a point-symmetric morphology.

\begin{figure*}
\begin{center}
\includegraphics[trim=2cm 1.5cm 1cm 2.0cm,scale=0.83]{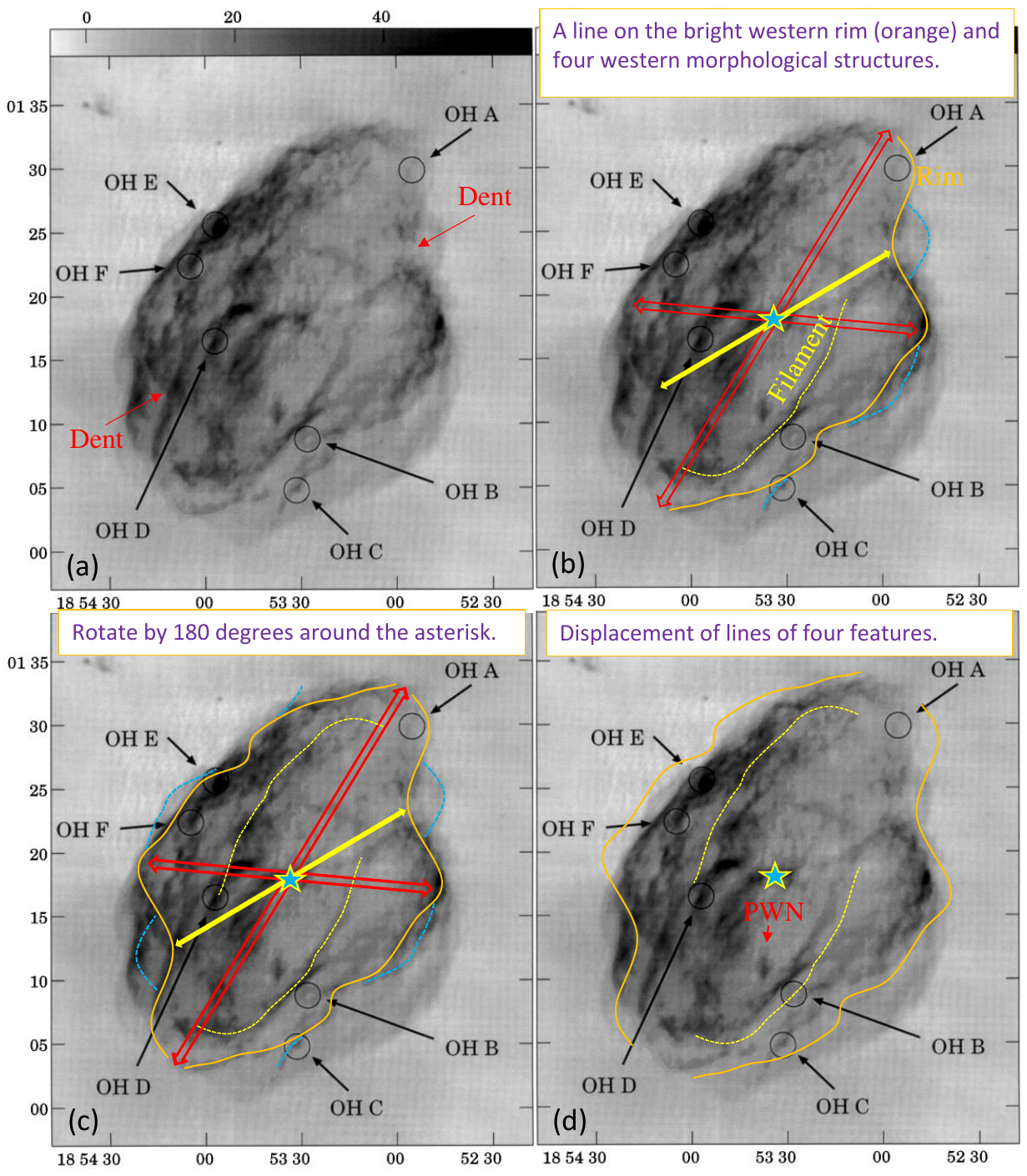} 
\caption{A radio image from \cite{Claussenetal1997}. The circles mark strong OH (1720 MHz) emission locations, which I do not refer to in this study. (a) The original image from \cite{Claussenetal1997}. I added two arrows that point at the two dents. (b) My identification of the bright western rim by the orange line. I also mark an inner filament (dotted-yellow line) and three structural features outside the western rim (pale-blue-dashed lines). The two double-line-double-headed red arrows are the directions of the two energetic jets that I propose to have shaped W44 during the explosion process. A third pair of jets shaped the inner filaments. (c) I copied the structural features I marked on panel (b) and rotated by $180^\circ$ around the point marked by an asterisk. This shows the point-symmetric part of W44. (d) I displaced outward the marks of the two rims and the two inner filaments to allow visual comparison of the observations with my marks. A short red arrow points at the PWN (pulsar wind nebula). Axes are right ascension and declination (B1950).  
}
\label{fig:Radio}
\end{center}
\end{figure*}

I found the radio image from \cite{Claussenetal1997} to fulfill my goals; I present this image in Figure \ref{fig:Radio} (the same image in the four panels). They mark the location of six regions with strong OH (1720 MHz) emission on that image, which I do not refer to in this study. On panel (a) of Figure \ref{fig:Radio}, I added two red arrows pointing at the structural features I term dent where the bright rim sharply bends inward. 

In panel (b) of Figure \ref{fig:Radio}, I mark several western structural features. I draw a line (orange) on the bright western rim of W44. I mark three structural features outside and attached to the rim on the west (dashed pale-blue lines). I draw a dotted-yellow line on an inner filament on the southwest. The yellow double-headed arrow points at the two dents. One double-line-double-headed red arrow is along the long dimension of W44 through the center of the point symmetrical structure that I mark with an asterisk; the other goes through the center and points at two opposite protrusions of the bright rim. The centers of the three double-headed arrows are at the asterisk. Panel (c) presents all the structural marks of panel (b), with their $180^\circ$-rotational symmetric marks. Namely, I took all structural marks from panel (b) and rotated them by $180^\circ$ around the asterisk. Panel (c) shows the point symmetrical structure that I identify in W44, primarily the two bright rims and the two inner filaments.

To allow a better visual inspection of the rim on the west and east and the two inner filaments, on panel (d) of Figure \ref{fig:Radio}, I displaced outward the marks of the two rims and the two inner filaments. I also mark the location of the pulsar-wind nebula (PWN; e.g., \citealt{Frailetal1996, Petreetal2002}). This PWN likely influenced the filament on the southwest; hence, I do not expect a perfect point-symmetry between the two opposite filaments marked by the dotted-yellow lines in Panel (b-d) of Figure \ref{fig:Radio}.  

\begin{figure}
\begin{center}
\includegraphics[trim=2cm 17.5cm 6cm 2.5cm,scale=0.9]{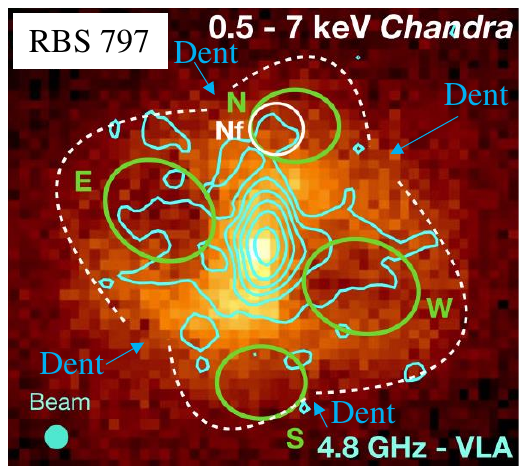} 
\caption{An X-ray image of the cooling flow cluster of galaxies RBS 797 from \cite{Ubertosietal2021}. Ellipses show the shape of the X-ray cavities. 
Cyan contours are radio emission at 4.8 GHz from \cite{Gittietal2013}. 
White dashed arcs encompassing the rims of the bubbles are from their image. 
Where two rims meet, there is a dent, as I indicate by the four arrows. 
}
\label{fig:Cluster}
\end{center}
\end{figure}

I suggest that each of the two dents is the zone where two jet-inflated structures (bubble/lobes/ears) touch each other on the bright rim. Similar dents are observed between jet-inflated bubbles in some planetary nebulae and clusters of galaxies. In Figure \ref{fig:Cluster} I present an image of the cooling flow cluster RBS 797 from 
\cite{Ubertosietal2021}. They mark (white dashed lines) the rims of the four jet-inflated bubbles, i.e., inflated by two pairs of jets.
Two adjacent rims have different directions where they meet, forming the dent. The similar structure of the dents in W44 with those in RBS 797, which are known to be inflated by jets, supports the jet-shaped modeling. In the case of the CCSNR W44, more than two pairs of jets have been involved. For example, I suggest that a third pair of jets shape the two opposite filaments. 

\section{Answering possible counterarguments}
\label{sec:Counter}
Let me answer some possible arguments against my interpretation of a jet-shaped point-symmetrical structure. 

(1) \textit{Counterargument:} There are many filaments and clumps in the radio image of W44, and one can find a point-symmetrical structure just by chance. 
\newline 
\textit{Answer:} It is hard to quantify the probability of a chance occurrence of the point symmetry I identified. I estimate it to be very low because the point symmetric structure does not consist of only a few clumps but rather two opposite rims along the long axis of the CCSNR and two inner filaments. As I indicated above, the point symmetry is not expected to be perfect because of several smearing processes. I can state that many researchers in other fields of astrophysics, such as clusters of galaxies and planetary nebulae, attribute such structures to pairs of jets. I consider my identification of a point-symmetrical structure to be robust.  

(2) \textit{Counterargument:}  The IR and X-ray images do not show a point symmetrical structure, as marked on the radio images. 
\newline 
\textit{Answer:} The IR image (Figure \ref{fig:W44GrichenerSoker}) shows many of the structures of the radio image (Figure \ref{fig:Radio}), but not all structures. For example, the IR shows part of the rims but is missing other parts, like the east dent. This situation is similar to other point-symmetrical morphologies of CCSNe, e.g., Cassiopeia A's IR and X-ray images reveal point-symmetrical morphologies but do not fully overlap \citep{BearSoker2025}. The X-ray map of W44 shows the inner part, so it does not show the point symmetry of the outer rims. It does follow an S-shape morphology \citep{GrichenerSoker2017}, a point-symmetrical morphology type, but it seems that the X-ray emitting gas fills the space inside the rims and filaments.    

(3) \textit{Counterargument:} Interaction with the cloud surrounding W44 shaped its morphology. 
\newline 
\textit{Answer:}  In general, interaction with ambient gas cannot explain the properties of the point-symmetrical structure of CCSNRs \citep{SokerShishkin2024Vela}. In the case of W44, the protrusions from the dents (pale-blue-dashed lines) imply that no dense ambient clumps form the dents. Other protrusions outside the rims on both sides show that it is not the ambient medium that shapes the point-symmetrical morphology. The ambient medium can shape these protrusions; hence, the protrusiona are not equal on both sides, i.e., the point symmetry is not perfect. Furthermore, the ambient gas cannot shape the two inner filaments part of the point-symmetrical morphology. 

(4) \textit{Counterargument:} Post-explosion jets shaped the point-symmetrical structure: 
\newline 
\textit{Answer:} The Point symmetrical structural features involve the entire CCSNR with an elongated structure. Hence, the energy of the shaping jets must be a large fraction of the explosion energy. It is unlikely that post-explosion jets carry such energy, and it is not clear why post-explosion jets should jitter. 

\section{Summary}
\label{sec:Summary}

I examined the complicated structure of CCSNR W44 (figures \ref{fig:colored} and \ref{fig:W44GrichenerSoker}) by marking some structural features on the west side of a radio image of W44 (panel b of figure \ref{fig:Radio}), and then rotate it around a central point (the asterisk). The rotated markings overlap with similar structures on the east side (panels c and d of Figure \ref{fig:Radio}), signify the point-symmetrical morphology of W44.
I attribute the shaping of the rims (orange lines) to two pairs of jets (red double-headed arrows). A third pair might have shaped the inner filaments. The dents are the zones where two jet-inflated structures meet. Such jet-shaped dents are observed in planetary nebulae and clusters of galaxies, one of which I present in figure \ref{fig:Cluster}.  

In Section \ref{sec:Counter}, I presented answers to some possible counterarguments to my claims. Mainly, I explained why the cloud that CCSNR W44 interacts with cannot account for the point-symmetrical morphology. This interaction rather smear point-symmetrical structures. 

The result of this short study, which identifies a point symmetrical morphology in CCSNR W44, is of great importance to the JJEM. According to the JJEM, stochastic pairs of jets explode most (possibly all) CCSNe, implying an explosion with a point-symmetric morphology. However, most pairs of jets are choked in the core, leaving no observable imprints on the CCSNR morphology. Instabilities during the explosion process, the NS natal kick and a PWN when existing, and interaction with the circumstellar material and the interstellar medium smear point-symmetrical structures that the shaping pairs of jets form. 
These smearing processes lead to imperfect point-symmetrical morphologies and can even completely erase the point-symmetric structures.  

For that, any CCSNR that reveals a point-symmetrical structure is a treasure to the development of the JJEM. SNR W44 is one of only 12 CCSNRs with identified point-symmetrical morphology \citep{Soker2025Two}. 
This study shows how a new analysis might reveal a different morphology, motivating more profound studies of CCSNRs for point-symmetrical morphologies.

This study's new identification of a point-symmetrical CCSNR, W44, adds to the growing evidence that the JJEM is the primary explosion process of CCSNe. 

\section*{Acknowledgements}
A grant from the Pazy Foundation supported this research.


\end{document}